\documentclass[footinbib,a4paper,aps,superscriptaddress,reprint,twocolumn,preprintnumbers,amsmath,amssymb,nobalancelastpage,10pt,longbibliography]{revtex4-1}
\usepackage[english]{babel}

\usepackage{color,amsthm,amsmath,amsxtra,amssymb , amsfonts,dsfont,graphicx,bm,bbold}

\usepackage{amsmath}
\usepackage{verbatim}
\usepackage{graphicx}
\usepackage{color}
\usepackage{tikz}
\usepackage{siunitx}
\usepackage{physics}
\usepackage{subfigure}
\usepackage{float}

\usepackage{mathptmx}
\usepackage{amssymb}
\usepackage{amsmath}
\usepackage{amsfonts}
\usepackage{array}

\usepackage{bm}
\usepackage{xcolor}
\usepackage{braket}

\definecolor{mygrey}{gray}{0.35}
\definecolor{myblue}{rgb}{0.2,0.2,0.8}
\definecolor{myzard}{cmyk}{0,0,0.05,0}
\definecolor{mywhite}{rgb}{1,1,1}
\definecolor{myred}{rgb}{1,0.,0.3}

\usepackage[colorlinks=true,citecolor=myblue,linkcolor=myred]{hyperref}

\def\beq{\begin{equation}}
\def\eeq{\end{equation}}

\newcommand{\st}[1]{_\text{#1}}

\newcommand\jj{\mathbf{j}}

\newcommand\kk{\mathbf{k}}

\newcommand\rr{\mathbf{r}}

\newcommand\bh{\hat b}

\newcommand{\mean}[1]{\langle #1\rangle}

\newcommand{\pa}[1]{\left( #1\right)}

\def\multiset#1#2{\ensuremath{\left(\kern-.3em\left(\genfrac{}{}{0pt}{}{#1}{#2}\right)\kern-.3em\right)}}

\usepackage{accents}

\usepackage{hyperref}

\definecolor{mygrey}{gray}{0.35}
\definecolor{myblue}{rgb}{0.2,0.2,0.8}
\definecolor{myzard}{cmyk}{0,0,0.05,0}
\definecolor{mywhite}{rgb}{1,1,1}
\definecolor{mywhite}{rgb}{1,1,1}
\definecolor{myred}{rgb}{1,0.,0.3}

\newcommand{\co}[1]{\left[ #1\right]}

\begin{document}


\title{Tuning long-range fermion-mediated interactions in cold-atom quantum simulators}

\author{Javier Arg\"uello-Luengo}
\email{javier.arguello@icfo.eu}
\affiliation{ICFO - Institut de Ci\`encies Fot\`oniques, The Barcelona Institute of Science and Technology, Av. Carl Friedrich Gauss 3, 08860 Castelldefels (Barcelona), Spain}

\author{Alejandro Gonz\'alez-Tudela}
\email{a.gonzalez.tudela@csic.es}
\affiliation{Institute of Fundamental Physics IFF-CSIC, Calle Serrano 113b, 28006 Madrid, Spain.}

\author{Daniel Gonz\'{a}lez-Cuadra}\email{daniel.gonzalez-cuadra@uibk.ac.at}
\affiliation{Institute for Theoretical Physics, University of Innsbruck, 6020 Innsbruck, Austria}
\affiliation{Institute for Quantum Optics and Quantum Information of the Austrian Academy of Sciences, 6020 Innsbruck, Austria}

\begin{abstract}

Engineering long-range interactions in cold-atom quantum simulators can lead to exotic quantum many-body behavior. Fermionic atoms in ultracold atomic mixtures can act as mediators, giving rise to long-range RKKY-type interactions characterized by the dimensionality and density of the fermionic gas. Here, we propose several tuning knobs, accessible in current experimental platforms, that allow to further control the range and shape of the mediated interactions, extending the existing quantum simulation toolbox. In particular, we include an additional optical lattice for the fermionic mediator, as well as anisotropic traps to change its dimensionality in a continuous manner. This allows us to interpolate between power-law and exponential decays, introducing an effective cutoff for the interaction range, as well as to tune the relative interaction strengths at different distances. Finally, we show how our approach allows to investigate frustrated regimes that were not previously accessible, where symmetry-protected topological phases as well as chiral spin liquids emerge.
\end{abstract}

\maketitle

\emph{Introduction.-} Long-range interactions between ultra-cold atoms are known to be the source of exotic many-body phenomena, including supersolid~\cite{buchler2003, Scarola_2005, Leonard_2017, Tanzi_2019, Bottcher_2019, Chomaz_2019}, magnetic~\cite{Micheli_2006, Barnett_2006, Gorshkov_2011, Maik_2012, vanBijnen_2015, Glaetzle_2015} and topological phases~\cite{Manmana_2013, Yao_2013, Yao_2018, Leseleuc_2019, Samajdar_2021, Verresen_2021, Semeghini_2021, Gonzalez_2022, Fraxanet_2022,Bello2019a,Bello2022}, soliton trains~\cite{Karpiuk2004,Santhanam2006,DeSalvo2019,Cheiney2018}, quantum droplets~\cite{Ferrier-Barbut2016,Chomaz2016,Cabrera2018}, or roton spectra~\cite{mottl12a,Feng2019}. Besides, if the long-range interactions appear between fermionic atoms, they can be harnessed to build analogue simulators for quantum chemistry~\cite{Arguello-Luengo2019,Arguello-Luengo2020,Arguello-Luengo2021} and high-energy physics~\cite{Zohar_2015, Banuls_2020, Gonzalez_2020c, Aidelsburger_2022}. Unfortunately, these interactions do not appear naturally between neutral atoms, since they generally interact through (local) elastic collisions~\cite{bloch05a}. This is why finding ways to engineer and control effectively such long-range atomic interactions is one of the most pressing issues in atomic physics nowadays.

As it occurs in nature, a conventional way of obtaining long-range interactions is through the exchange of mediating particles. For example, the exchange of photons through the atomic optical transitions leads to dipolar interactions ($1/r^3$) in the case of highly magnetic~\cite{Lu2011,Giovanazzi2002,griesmaier05a} and Rydberg atoms~\cite{zeiher16a,Guardado-Sanchez2018,Lienhard2018,Browaeys2020}, or dipolar molecules~\cite{Ni2008,deiglmayr08a,Stuhler2005,Hazzard2014}. One can extend their range by shaping the photonic field with cavities~\cite{Baumann2010,Munstermann2000,mottl12a,ritsch13a,Vaidya2018,Vaidya2018} or nanophotonic structures~\cite{douglas15a,Gonzalez-Tudela2015b,Chang2018}. However, such photon-mediated interactions are accompanied by dissipation, which needs to be controlled to profit from them. An orthogonal direction that is recently being considered consists in using fermionic atoms in atomic mixtures as mediators~\cite{Chui2004,gunter06a,Sinha2009,Best2009,Mering2010,Polak2010,Ferrier-Barbut2014,De2014,Suchet2017,DeSalvo2017,Feng2019,DeSalvo2019,Edri2020,Zheng2021}. Such fermion-mediated interactions have been predicted~\cite{santamore2008,De2014,Suchet2017} to lead to the Ruderman-Kittel-Kasuya-Yosida (RKKY)-type interactions appearing in solids~\cite{Kasuya1956,ruderman54a,Yosida1957}, which have a power-law, oscillating nature, fixed by the dimensionality and density of the Fermi gas. With the recent experimental observation of such interactions~\cite{DeSalvo2019,Edri2020}, a timely question that has been scarcely explored~\cite{Feng2019} is how one can further tune those interactions to be able to explore new phenomenology with them.

\begin{figure}[t]
  \centering
  \includegraphics[width=1\linewidth]{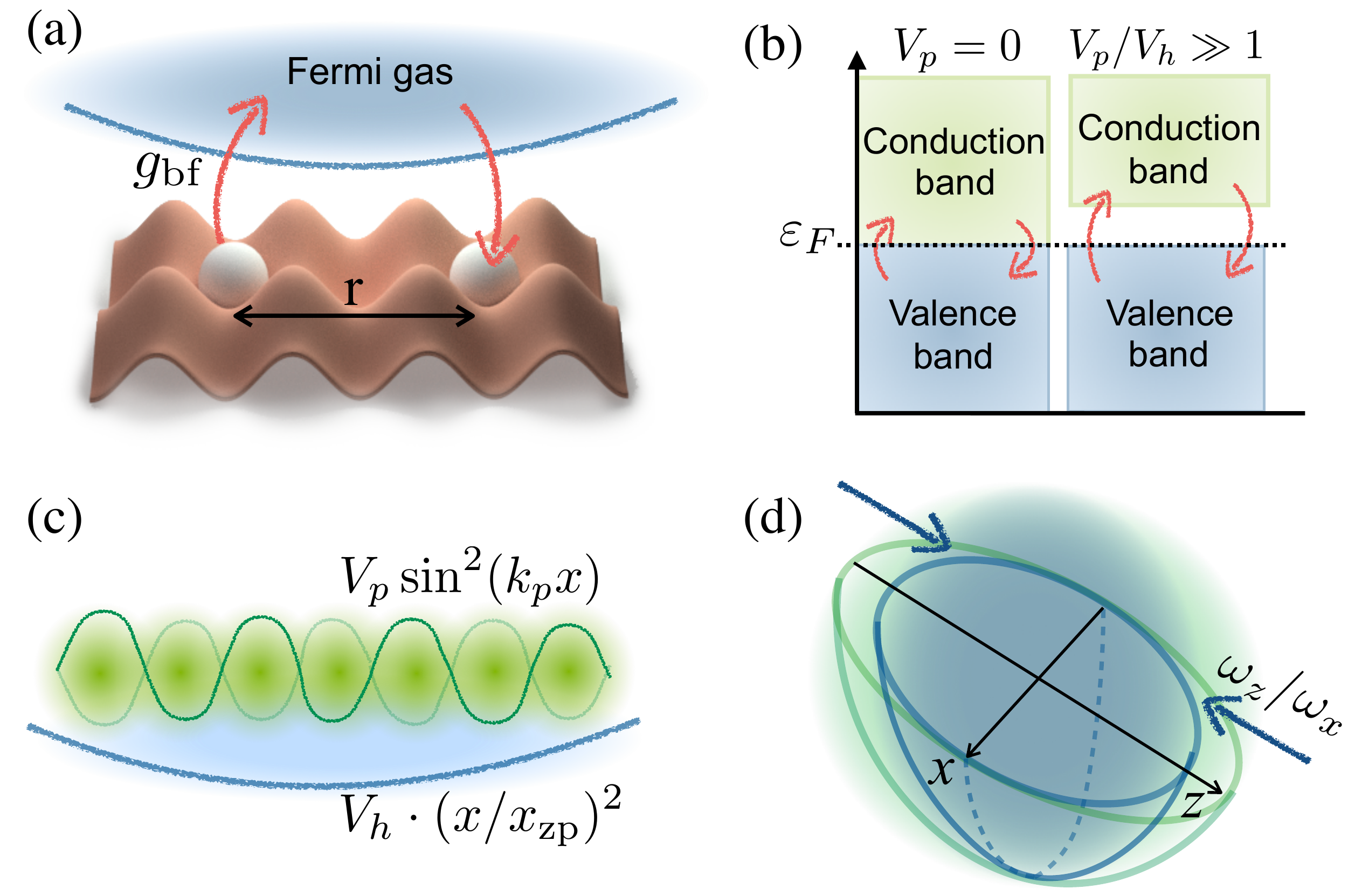}
\caption{\label{fig:mechanism} (a) Two bosonic atoms (white) separated by distance $r$ and trapped in an optical lattice (red) experience an effective long-range interaction mediated by a Fermi gas trapped in an harmonic potential (blue). (b) The contact Bose-Fermi interactions ($g_{\rm bf}$) virtually populates the conduction band of the Fermi gas. (c) An additional oscillatory potential induces a gap between the valence and conduction gap, exponentially damping the mediated interactions. (d) By controlling the strength of the trapping potential in an orthogonal direction, $\omega_z/\omega_x$, one can continuously tune the dimension of the Fermi gas from 1D to 2D, introducing additional modulating frequencies as will be shown in Fig.~\ref{fig:4}.}
\end{figure}

In this work, we take advantage of the flexibility offered by ultracold atomic platforms to control the range and shape of long-range fermion-mediated interactions, going beyond the conventional RKKY-type interactions encountered in solid-state systems. This allows us to design a quantum simulation toolbox that can be used to prepare, for instance, frustrated phases that are not accessible using other approaches. The paper is organized as follows. First, we review how to derive the effective fermion-mediated RKKY interactions for a Fermi-Bose mixture of ultracold atoms. We then introduce an additional optical potential for the Fermi gas and show how the range of the interactions can be interpolated from a power law to an exponential decay by tuning the ratio between the periodic potential and the confining harmonic trap. This allows at the same time to select the ratios between interactions at different distances within a non-vanishing range. We then show how, for a hardcore bosonic chain immersed in the fermionic cloud, the resulting interactions can be used to prepare frustrated phases with non-trivial topological properties. Finally, we explore an extra tuning knob by continuously changing the dimensionality of the cloud using different harmonic frequencies in each spatial direction. This control allows one to further tune the interaction ratios which, as we indicate, might be useful to obtain 2D chiral spin liquid phases~\cite{Zhu2016}.

\emph{Effective fermion-mediated interactions.-} Let us consider an atomic mixture as depicted in Fig.~\ref{fig:mechanism}(a), where one species corresponds to non-interacting spinless fermions trapped by an harmonic potential,
\begin{equation}
\label{eq:harmonic_potential}
    V(x,y,z)= \frac{m_f}{2}\co{\omega_x^2 x^2+\omega_y^2 y^2 +\omega_z^2 z^2}.
\end{equation}
The fermionic Hamiltonian reads $\hat{H}_f= \sum_n \varepsilon _n \hat{c}^\dagger_n \hat{c}_n$, where $\hat{c}^{(\dagger)}_n$ is the (creation) annihilation operators associated with the $n$-th eigenstate of this oscillator, with energy $\varepsilon _n$. For concreteness, and without lack of generality, we assume that the other atomic species is bosonic and  it is trapped in an optical lattice, $V\st{lat}(\rr)$, with fixed lattice spacing $d$. 
While Pauli blocking prevents s-wave interactions among spin-polarized fermion, one can still account for boson-boson and boson-fermion collisions. This leads to a bosonic Hamiltonian of the form,
\begin{equation}
\begin{split}
    \hat H\st{b}&=\int d\rr\, \phi^\dagger(\rr) \pa{\frac{-\hbar^2}{2m\st{b}}\nabla^2+V\st{lat}(\rr)}\phi(\rr)\\
    & +g\st{bb}\int d\rr \phi^\dagger(\rr) \phi^\dagger(\rr)\phi(\rr)\phi(\rr)\,,
    \end{split}
\end{equation}
and a Bose-Fermi density-density interaction,
\begin{align}
    \hat H\st{I}=g\st{bf}\int d\rr \phi^\dagger(\rr) \psi^\dagger(\rr)\phi(\rr)\psi(\rr) \
\end{align}
where the field operator $\phi(\rr)\,[\psi(\rr)]$ describes the anihilation of a boson [fermion] and $g\st{bb} (g\st{bf})$ is the bosonic (interspecies) coupling constant, which is experimentally tunable through magnetic Feshbach resonances~\cite{chin10a}.

To obtain the effective fermion-mediated interaction, we assume that i) the fermionic timescales are much faster than the bosonic ones, $t_b,U_b\ll \hbar \omega_{x,y,z}$, ii) and the $N$ fermionic atoms are occupying all states up to the Fermi energy, $\varepsilon_F=\varepsilon_N$, so that the state can then be written as $\ket{\Omega} = \prod_{n=1}^N\hat{c}^\dagger_n\ket{0}$. With these assumptions, and taking $\hat{H}_0=\hat{H}_b+\hat{H}_f$ as the unperturbed Hamiltonian and $\hat{H}_I$ as a perturbation, we obtain an effective potential for the bosons~\cite{De2014},
\begin{equation}
\label{eq:perturbation}
\hat H\st{eff} = \varepsilon_F+\hat H_b+ G\iint d\rr d\rr' F_{\rr,\rr'} \phi^\dagger(\rr) \phi^\dagger(\rr')\phi(\rr)\phi(\rr') \,,
\end{equation}
where $G=2m_f g\st{bf}^2/(\hbar^2 x\st{zp}^4)$ and $x\st{zp} = [\hbar/(2m\omega_x)]^{1/2}$. Here, the last term arises from the second-order perturbation $\sum_{m \neq i} \abs{\bra{m} \hat{H}_I \ket{i}}^2/(E_m - \varepsilon_F)$, where the initial state $\ket{i} = \ket{\Omega} \ket{\{r_{\rm b}\}}$ belongs to the ground-state manifold of $\hat{H}_0$ for bosonic atoms placed in positions $\{r_{\rm b}\}$, and $\ket{m} = \hat{c}^\dagger_{m} \hat{c}^{\vphantom{\dagger}}_{n}\ket{\Omega} \ket{\{r_{\rm b}\}}$ is a particle-hole excited state outside the manifold, with energy $E_m$. 
Note that due to the conservation of fermionic parity, fermions needs to be exchanged twice to generate a potential, unlike photons which can be exchanged only once. This has important implications for the sign and shape of $F_{\rr,\rr'}$~\cite{Jiang2021}. Interestingly, this mediated potential only depends on the bosonic separation $F_{\rr,\rr'}\approx F\pa{\abs{\rr-\rr'}}$.

Since it will be useful to interpret the results of this manuscript, let us review here the properties of an untrapped free Fermi gas with energy dispersion $\varepsilon^0_\kk=\frac{\hbar^2|\kk|^2}{2m_f}$, which provides a first approximation in the limit $N\gg 1$.
Analytical expressions in this limit can be found in all spatial dimensions~\cite{Rusin2017,Fischer1975,ruderman54a}. For example, in the one-dimensional case ($N\omega_x\ll \omega_{y,z}$), $F(r)$ expands in the limit $k_{\rm F} r\gg 1$ as~\cite{Rusin2017}:
\begin{align}
\label{eq:fr1D}
        F\st{1D}(r)\propto  \frac{-1}{k_{\rm F} r}\pa{\cos\pa{2k_{\rm F} r}+\frac{\sin\pa{2k_{\rm F} r}}{2k_{\rm F} r}}\,,
\end{align}
whereas in the two-dimensional case ($\omega_x=\omega_z\ll \omega_{y}/N$), it expands as~\cite{Fischer1975}:
\begin{align}
\label{eq:fr2D}
    F\st{2D}(\rr) \propto  -\frac{x\st{zp}^2}{r^2}\pa{\sin(2k_{\rm F} r)-\frac{\cos(2k_{\rm F} r)}{4k_{\rm F} r}}\,.
\end{align}

In both dimensions, the interactions share some common features: i) the fermion-mediated interactions are attractive in the limit $r\rightarrow 0$, regardless of the sign of $g_{bf}$. The mean-field intuition is that, for $g_{bf}>(<)0$, the Fermi gas tends to avoid (be attracted to) the bosons, reducing (increasing) the fermionic density, and thus, the bosons feel more attracted to this place. ii) Asymptotically, they lead to longer range interactions $\sim r^{-1}$ (1D) and $\sim r^{-2}$ (2D) than dipolar ones ($\sim r^{-3}$). iii) The interactions oscillate with an effective length inversely proportional to the Fermi momentum, $k_{\rm F}=\sqrt{2m_f\varepsilon\st{F}}$. Thus, choosing the wavevector of the bosonic optical lattice potential $k_L$, one can induce (anti-)ferromagnetic interactions if $k_L=(2)k_{\rm F}$, or incommensurate ones ($k_L/k_{\rm F} \in \mathds{I}$).

For a sufficiently deep optical lattice for the bosons, only its lowest motional bands get populated. Wannier functions $w_\jj(\rr)$ centered at the lattice sites $\jj$ become a convenient description for the bosonic fields, $\phi(\rr)=\sum_\jj w_\jj(\rr) \bh_\jj$. Projecting in this basis the effective Hamiltonian~\eqref{eq:perturbation}, one obtains an extended Bose-Hubbard model,

\begin{align}
\hat{H}\st{eff}=- t\st{b} \sum_{\mean{\jj,\jj'}}\hat{b}^\dagger_{\jj} \hat{b}^{\vphantom{\dagger}}_{\jj'} + \frac{U\st{b}}{2} \sum_\jj \hat{n}^{\rm b}_{\jj} (\hat{n}^{\rm b}_{\jj} - 1)+\sum_{\jj,\jj'} \upsilon_{\jj,\jj'}\hat{n}^b_\jj \hat{n}^b_{\jj'}\,,
\end{align}
with nearest-neighbor tunneling strength $t$ and on-site interaction $U\st{b}= g\st{bb} \int d\rr \abs{w_{\jj}(\rr)}^4$. Here $\hat{b}^{(\dagger)}_{\jj}$ are the (creation) annihilation operators of a bosonic atom on site $\jj$, $\hat{n}^{\rm b}_\jj=\hat{b}^\dagger_\jj \hat{b}^{\vphantom{\dagger}}_\jj$ is the bosonic number operator, and the effective potential $\upsilon_{\jj,\jj'}$ terms can be obtained from the perturbed potential~\eqref{eq:perturbation} as

\begin{align}
    \upsilon_{\jj,\jj'} &= G \iint d\rr d\rr'\, F\pa{|\rr-\rr'|}\abs{w_{\jj}(\rr)}^2 \abs{w_{\jj'}(\rr')}^2\,.
\end{align}

\begin{figure}[t]
  \centering
  \includegraphics[width=1.0\linewidth]{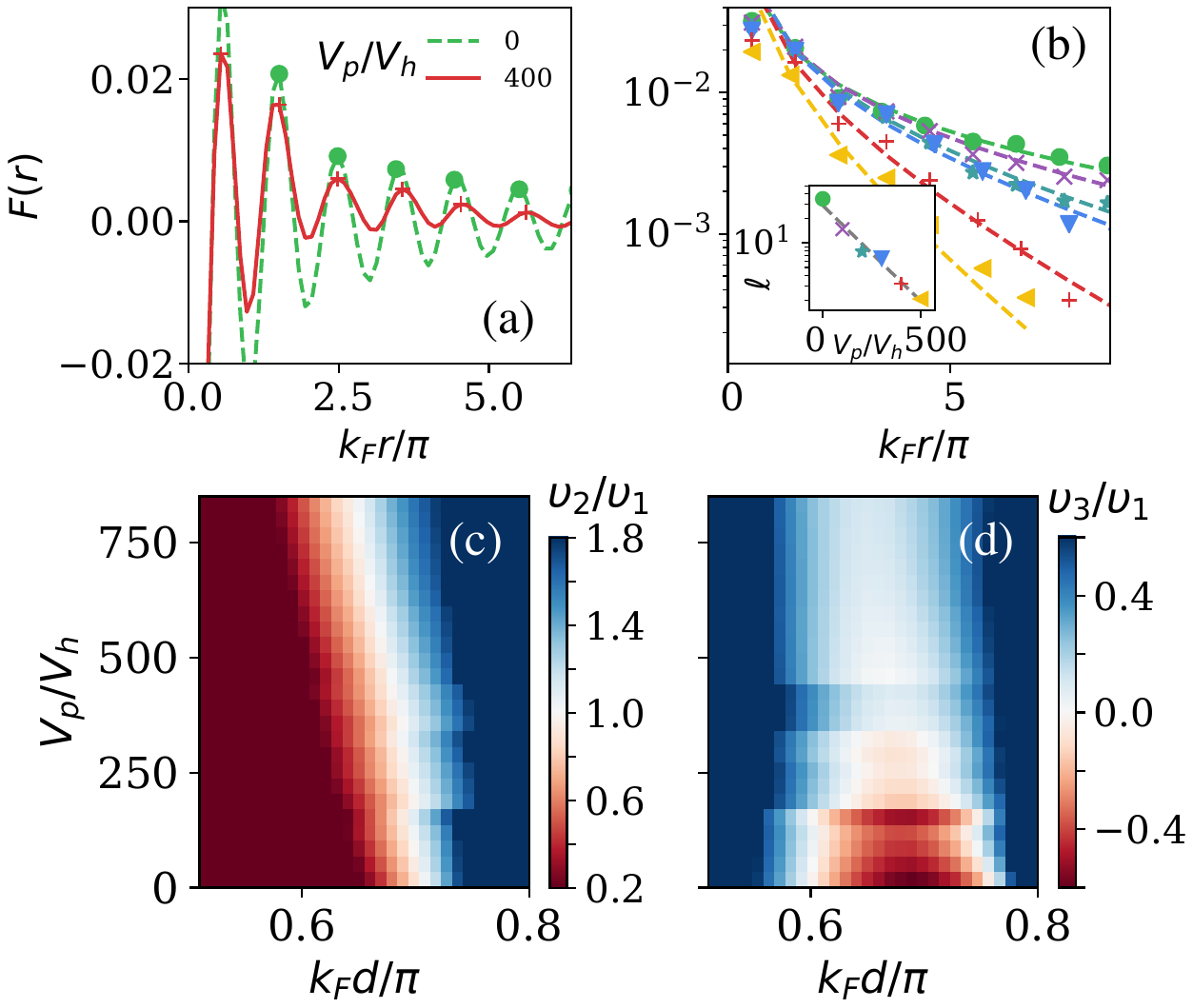}
\caption{(a) Effective potential between two atoms in one dimension mediated by a Fermi gas trapped in an harmonic trap with (green) and without (red) an extra periodic trap of strength $V_p/V_h=400$. Markers indicate the oscillation maxima. (b) Value of the maxima for increasing values of $V_p/V_h$. The inset shows the decay length $\ell$ of a fitted Yukawa interaction, $\sim \text{exp}[-k_Fr/(\pi\ell)]/r$. (c-d) Strength of the second $\upsilon_2 / \upsilon_1$ (c) and third $\upsilon_3 / \upsilon_1$ (d) neighbour interactions as a function of $V_p/V_h$ and the effective lattice spacing $k_{\rm F}d$. Here, the bosonic Wannier function $w_\jj(\rr)$ is approximated by a gaussian distribution with, $x\st{width}/d=0.17$, consistent with a lattice wavelength $\lambda=784.7$ nm for $^{87}$Rb~\cite{Hruby2018} and trapping depth $s=12$. We also took $N=200$, $k_p x\st{zp}=13.2$, and scattering length $a\st{bf}=25a_0$.}\label{fig:2}
\end{figure}

\emph{Controlling the range of the interactions.-} We now show how the range of the effective interactions can be controlled by adding a periodic potential $V_p\sin^2(k_p x)$ to the previous fermionic trap in the 1D case, $V(x) = V_h\cdot (x/x_{\rm zp})^2$, where $V_h=\hbar \omega_x/4$, as illustrated in Fig.~\ref{fig:mechanism}(c). In the following, we fix the number of fermionic atoms, while we vary the ratio $V_p/V_h$ by modifying the depth of the periodic potential. It is expected that a value of $V_p\neq 0$ opens up a gap in the energy dispersion of the fermionic excitations, introducing a cutoff in the interaction range if the Fermi energy lies within the band gap (see Fig.~\ref{fig:mechanism}(b)). This is guaranteed by choosing an appropriate wavevector $k_p$ for the fermionic optical lattice, such that $k_p x\st{zp}\sim \sqrt{N}$ and the oscillatory potential maximally hybridizes with the $N-1$ nodes of the highest occupied state.

In Fig.~\ref{fig:2}(a), we show the fermion-mediated interaction appearing in a 1D Fermi gas for two values of $V_p/V_h$ corresponding to the pure harmonic case ($V_p/V_h=0$, dashed green) and a ratio $V_p/V_h=400$ (solid red). We observe how the periodic potential tends to (exponentially) cut the range of the interaction, inducing a purely positive potential for distances $k_{\rm F} r>2$. In Fig.~\ref{fig:2}(b), we plot the maximum relative values of the fermion-mediated interactions at the oscillations for increasing values of $V_p/V_h$, where it is more evident the transition from a power-law decay for small $V_p/V_h$, to an exponentially decaying Yukawa-like interaction when $V_p/V_h$ is large. We can therefore control the effective interaction range, given by the exponential decay length $\ell$, which is exponentially reduced by $V_p/V_h$, as $\ell \sim e^{-\alpha V_p/V_h}$ (inset).

Besides, playing with the effective lattice separation of the bosonic species $k_{\rm F} d$ (which can be controlled through the frequency of the harmonic trap, adjusting $k_p$ accordingly), one can identify diverse choices of induced interactions, as we illustrate in Figs.~\ref{fig:2}~(c-d). There, we see for example that there are regions for $V_p/V_h$ and $k_{\rm F} d$ (coloured in white) where the potential for the nearest and next-nearest neighbours coincide $\upsilon_2\approx\upsilon_1$, while interactions among longer distance atoms are very weak, $\upsilon_3/\upsilon_1\approx 0$. This is important, as such potentials can be the source of frustrated quantum many-body phases.

\begin{figure}[t]
  \centering
  \includegraphics[width=1.0\linewidth]{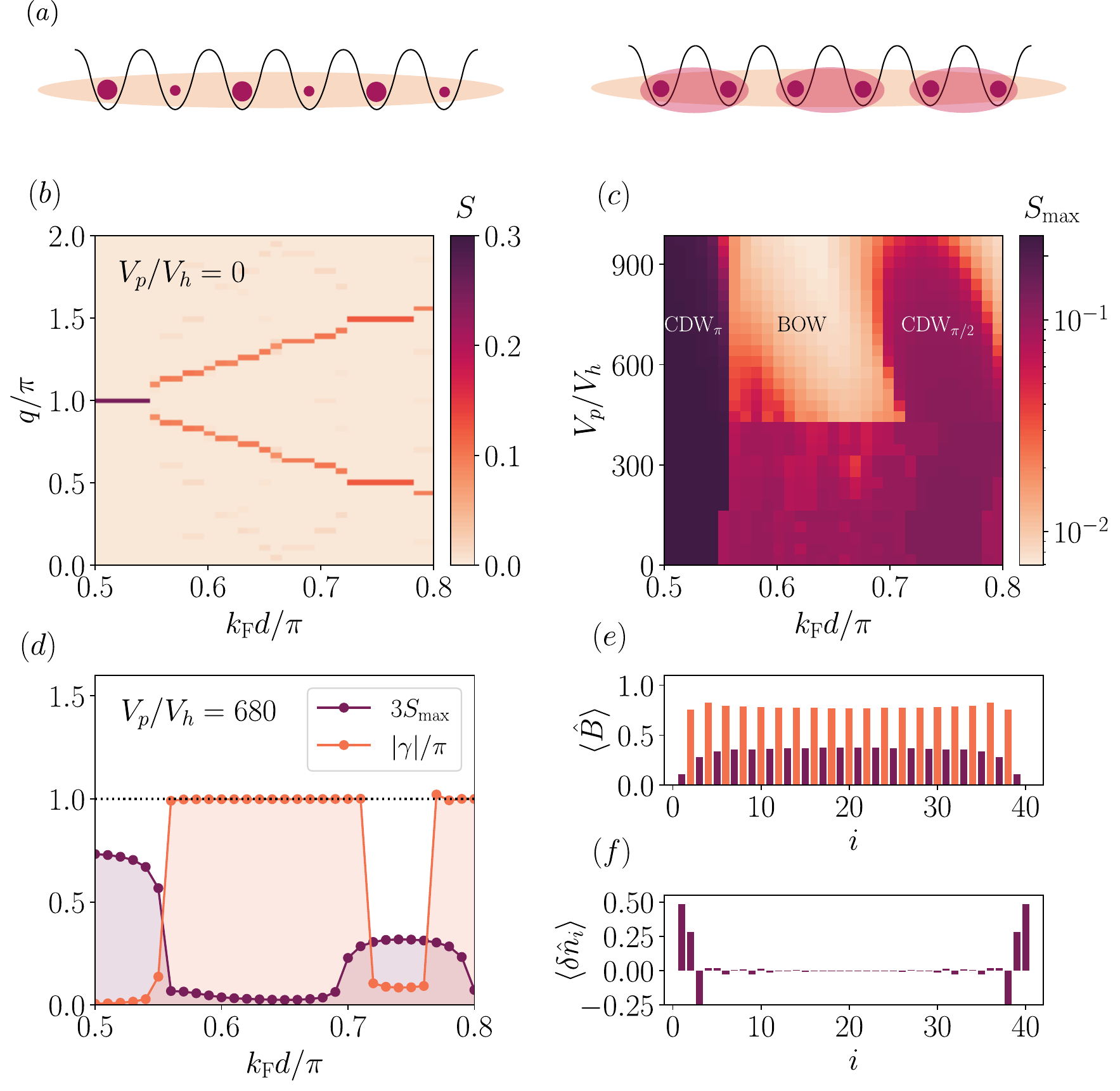}
\caption{\label{fig:3} (a) Real-space configuration of the CDW$_{\pi}$ (left), where spheres of different sizes correspond to an alternating atomic occupation, and frustrated BOW (right) with a dimerized bond structure. (b) $S(q)$ as a function of $k_{\rm F} d$ for a chain with $L = 60$ sites and bosonic density $\rho = 1/2$, for $V_p = 0$. The phase diagram presents a staircase structure, where every step corresponds to a CDW$_q$ characterized by a peak in $S(q)$. (c) Value of $S$ at the peak, $S_\text{max}$ as a function of $k_{\rm F}d$ and $V_p / V_h$, showing how the CDW orders melt for sufficiently large values of $V_p / V_h$, giving rise to a frustrated BOW. (d) $S_\text{max}$ and Berry phase $\gamma$ as a function of $k_{\rm F}d$ for $V_p / V_h = 680$, showing the non-trivial topological nature of the BOW phase. Real-space configuration of the BOW phase for $L = 40$, showing (e) dimerized bonds and (f) localized edge-states at the boundaries. Parameters as in Fig.~\ref{fig:2}.}
\end{figure}

\emph{Quantum simulation toolbox for frustrated phases.-} To illustrate the last point, we analyze the phase diagram of a 1D chain of hardcore bosons ($U_b/t_b \rightarrow \infty$) whose interactions are mediated by a 1D Fermi gas. The different phases can be distinguished using the structure factor,
\begin{equation}
    S(q) = \frac{1}{L^2}\sum_{i, j}\langle \delta \hat{n}^\text{b}_i \delta \hat{n}^\text{b}_j \rangle e^{i q (r_i-r_j)},
\end{equation}
where $\delta \hat{n}^\text{b}_i = \hat{n}^\text{b}_i - \rho$ and $\rho = 1 / L \sum_i \langle \hat{n}^\text{b}_i \rangle$ is the bosonic density.

Using a density-matrix renormalization group (DMRG) algorithm~\cite{Hauschild_2018} with fixed bond dimension $D = 200$, we calculate the ground state of a periodic chain with $L = 60$ sites and half-occupation, $\rho = 1/2$. Fig.~\ref{fig:3}(b) shows how, for $V_p / V_h = 0$, $S(q)$ develops a clear peak at a certain value $q_0$ that varies with $k_{\rm F}d$. The value at the peak $S_\text{max} = S(q_0)$ can be used as an order parameter, revealing in this case a staircase structure where every step corresponds to a charge density wave (CDW) phase with long-range order in the atomic density. For each of them, the order is characterized by the momentum $q_0$, and we labelled these phases as CDW$_{q}$. As an example, we depict in Fig.~\ref{fig:3}(a) the real-space density for CDW$_{\pi}$.

In Fig.~\ref{fig:3}(c), we can observe how the situation changes as we increase the value of $V_p / V_h$. If the amplitude of the periodic potential is sufficiently large, a disordered phase emerges between the different CDW$_q$ phases, where $S(q)$ vanishes at all momenta. This is an example of a frustrated phase~\cite{Lacroix_2011}, where the density order melts due to quantum fluctuations enhanced by competing interactions in a region where the different density orders are close in energy. Instead, a bond order develops (Fig.~\ref{fig:3}(a)), characterized by a non-zero value of the order parameter $B=1/L\sum_\jj (-1)^j\langle \hat{B}\rangle$, with $\hat{B} = \hat{b}^\dagger_\jj \hat{b}^{\vphantom{\dagger}}_{\jj+1} + \text{H.c.}$ (Fig.~\ref{fig:3}(e)). This bond-order wave (BOW) is a strongly-correlated phase that can not be accessed through the conventional RKKY interactions~\cite{Mering2010}, since it requires comparable nearest and next-nearest neighbor interactions, while further-range interactions should vanish. This allows, in particular, CDW$_\pi$ and  CDW$_{\pi/2}$ to compete without allowing for other CDW$_q$ orders, such as the bond order is preferred. While this situation can be achieved for spinfull fermions with dipolar interactions~\cite{Julia-Farre_2021}, spinless particles require $\upsilon_2 / \upsilon_1 \approx 0.5$~\cite{Mishra_2011}, which is achieved here by varying the periodic potential, as shown previously. Similarly to the fermionic case~\cite{Julia-Farre_2021}, here the BOW phase possesses non-trivial topological properties. These are characterized by both a non-zero quantized value of the Berry phase~\cite{Hatsugai_2006} (Fig.~\ref{fig:3}(d)), calculated here through from the entanglement spectrum as explained in Ref.~\cite{Zaletel_2014}, and the emergence of localized protected states at the boundaries (Fig.~\ref{fig:3}(f)). We note that similar topological effects are observed in non-frustrated BOW phases induced instead by dynamical optical lattices~\cite{gonzalez_18,gonzalez_19a,gonzalez_19b,gonzalez_20a,gonzalez_20b, Chanda_2021, Chanda_2022}.

\emph{Controlling the shape of the interactions.-} Let us finally provide a way of tuning the interactions which is unique of atomic systems, enabled by the possibility to control the effective dimensionality of the fermionic gas. In particular, by superimposing three independent standing-wave potentials, each $\omega_i$ in Eq.~\eqref{eq:harmonic_potential} can be controlled independently for the three orthogonal directions by modifying their intensity (see Fig.~\ref{fig:mechanism}(d)). Starting from $\omega_{x}=\omega_y=\omega_{z}$ and increasing $\omega_y$, one can go smoothly from a 3D fermionic gas to an effective 2D one for $\omega_{x,z}\ll \omega_y/N$~\cite{Dyke2011,Martiyanov2010}. Similarly, increasing $\omega_z$ connects the 2D and the 1D case. Since the power-law exponent of $F(r)$ depends on the dimension $D$ as $1/r^D$, one expects that this method interpolates between different integer values.

\begin{figure}[tb]
  \centering
  \includegraphics[width=1.05\linewidth]{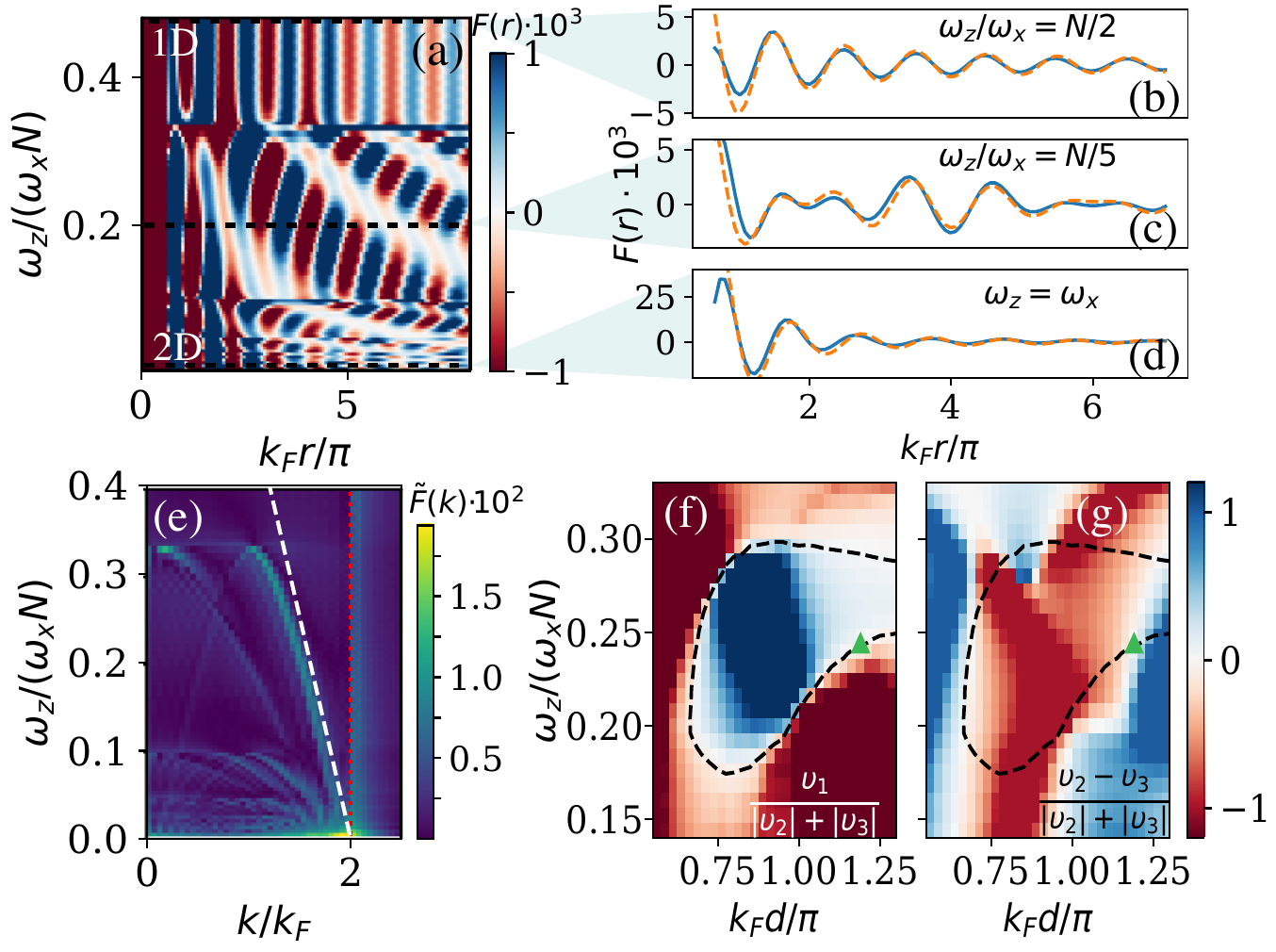}
  \caption{\label{fig:4} (a) Mediated potential $F(r)$ as a function of the anisotropy ratio $\omega_z/(N\omega_x)$ and effective atomic separation $k_{F}r$. Repulsive (attractive) forces are represented in red (blue). (b-d) Value of $F(r)$ for three anisotropy ratios (blue) indicated in (a) with black dashed lines, compared to the expected analytic results~(\ref{eq:fr1D},\ref{eq:fr2D}) (orange). (e) Cosine transform of  $F(r)$ as a function of $\omega_z/(N\omega_x)$, where the dotted lines correspond to the frequencies $2k_{\rm F}$ (red) and $\tilde{k}_1$ (white). (f,g) Relation between the nearest-neighbor potentials $\upsilon_{1,2,3}$ of a kagome lattice with lattice spacing $d$, as a function of $\omega_z/(\omega_xN)$ and $k_{F}d$. Dashed contour follows $\upsilon_1=0$. Here we took $N=250$ and the rest of the parameters as in Fig.~\ref{fig:2}.
  }
\end{figure}

We now explore the effect of this dimensional crossover in the effective interactions $F(r)$ for the two-to one-dimensional transition, while we maintain the bosons in 1D. Fig.~\ref{fig:4}(a) shows $F(r)$ as a function of the anisotropy ratio $\omega_z/(\omega_x N)$ and $k_{\rm F}d$, together with some cuts at the 1D / intermediate / 2D regimes in Figs.~\ref{fig:4}(b-d). Note that the dependence on $N$ is introduced because the crossover is expected in the limit $\omega_z/\omega_x\sim N$, where the energy of the highest-energy state is not enough to induce an excitation in the $z$-direction, and the interaction becomes effectively 1D. We observe that the interpolation is more intricate than initially expected. While in the limits $\omega_z/(\omega_x N)\gg (\ll) 1$ one recovers the expected 1D (2D) RKKY-type interactions, the intermediate dimensions acquire additional beating oscillations due to the presence of different harmonics in the potentials. This is more evident in Fig.~\ref{fig:4}(e), where we plot the corresponding cosine transform $\tilde F(k)$. The frequency $2k_{\rm F}$ appears in all intermediate dimensions and, through a careful analysis, we observe that additional frequencies appear associated to discrete values $\tilde k_n=2k_{\rm F}[1-n\omega_z/(\omega_xN)]$. The larger the value of $\omega_z/\omega_x$, the smaller is the contribution associated to smaller frequencies (as longer effective lengths that cannot fit the constrained direction vanish). In particular, in the range $\omega_z/(\omega_xN)\in (0.1,1/3)$, only contributions associated to $\tilde k_0$ and $\tilde k_1$ are dominant, leading to a smooth beating between the two frequencies in the potential, as we fit in Fig.~\ref{fig:4}(c) (dashed line).

Despite the apparent complexity of the fermion-mediated interactions within this dimensional crossover, they might lead to the appearance of novel many-body phases difficult to obtain otherwise. For example,e recent works have shown how chiral spin liquids can appear for hardcore bosons in kagome lattices with long-range interactions where the second and third neighbor terms are similar and the nearest neighbor interaction cancels, i.e., ($\upsilon_2\approx \upsilon_3$ and $\upsilon_1=0$)~\cite{Zhu2016}, a regime that is typically hard to access with conventional approaches. In Fig.~\ref{fig:4}(f,g), we make a search whether such regime would be accessible through this dimensional crossover, and find that indeed there are configurations where $\upsilon_1\approx 0$ (see contour line), while $\upsilon_2\approx \upsilon_3$ (green marker). Although further analysis is required, specially to account for the effect of further-range interactions in the phase diagram, our results show a promising avenue to investigate magnetic frustration and spin liquids states in 2D ultracold atomic mixtures using tunable long-range interactions.

\emph{Conclusions.-} We provide two strategies to control fermion-mediated interactions in ultracold atomic mixtures by modifying the fermionic confinement potential. First, we add an extra periodic potential to open a gap in the fermionic band. Then, we continuously modify the effective dimension of the fermionic gas by using anisotropic traps. In both cases, we characterize the emergent long-range interactions, obtaining a very versatile control over their range and shape. Finally, we consider different examples where this extended quantum simulation toolbox can lead to the exploration of frustrated quantum many-body phases that are not easily accessible with other approaches. Given the recent experiments in this direction~\cite{DeSalvo2019,Edri2020}, and the relatively simple tools that our proposal demands, we expect our results to guide near-future experiments on the topic.

\begin{acknowledgements}
 We are indebted to J. I. Cirac, V. Kasper and M. Lewenstein for very insightful discussions on the topic. JAL acknowledges support from 'la Caixa' Foundation (ID 100010434) through the fellowship LCF/BQ/ES18/11670016, Severo Ochoa Grant CEX2019-000910-S [MCIN/AEI/10.13039/501100011033], Generalitat de Catalunya (CERCA program), Fundació Cellex and Fundació Mir-Puig.
AGT acknowledges support from   CSIC Research   Platform   on   Quantum   Technologies   PTI-001, from  Spanish  project  PGC2018-094792-B-100(MCIU/AEI/FEDER, EU), and from the Proyecto Sinérgico CAM 2020 Y2020/TCS-6545 (NanoQuCo-CM). DGC is supported by the Simons Collaboration on Ultra-Quantum Matter, which is a grant from the
Simons Foundation (651440, P.Z.).
\end{acknowledgements}

%

\end{document}